\documentclass[prl,showpacs,twocolumn,amsmath,amssymb,floatfix,superscriptaddress]{revtex4-2}
\usepackage{amsmath}
\usepackage{amssymb}
\usepackage{graphicx}
\usepackage{bm}
\usepackage{color}
\usepackage{hyperref}
\usepackage{graphicx}
\usepackage{upgreek}
\usepackage{scalerel}
\usepackage{multirow}
 
\usepackage{pgffor}
\usepackage{pdfpages}
\usepackage{mathdots}
\usepackage{float}
\usepackage{silence}
\usepackage{physics} 
\usepackage{lscape}   
\usepackage{color, colortbl}
\usepackage[table]{xcolor}
\usepackage{comment}  
\makeatletter
\AtBeginDocument{\let\LS@rot\@undefined}
\makeatother

\begin{document}
\title{Non-Hermitian phase-biased Josephson junctions}
\author{Jorge Cayao}
\affiliation{Department of Physics and Astronomy, Uppsala University, Box 516, S-751 20 Uppsala, Sweden}
\author{Masatoshi Sato}
\affiliation{Center for Gravitational Physics and Quantum Information, Yukawa Institute for Theoretical Physics, Kyoto University, Kyoto 606-8502, Japan}

\date{\today} 
\begin{abstract}
We study non-Hermitian Josephson junctions formed by superconductors with a finite phase difference under non-Hermiticity, which naturally appears due to coupling to normal reservoirs. Depending on the structure of non-Hermiticity, captured in terms of retarded self-energies, the low-energy spectrum hosts topologically stable exceptional points either at zero or finite real energies as a function of the superconducting phase difference. Interestingly, the corresponding phase-biased supercurrents may acquire divergent profiles at such exceptional points. This instance is a natural and unique non-Hermitian effect that signals a possible way to enhance the sensitivity of Josephson junctions. Our work opens a way for realizing unique non-Hermitian phenomena due to the interplay between non-Hermitian topology and the Josephson effect.
\end{abstract}

\maketitle

Non-Hermitian (NH) systems have recently attracted enormous interest due to their potential for realizing unexpected states of matter that are impossible in the Hermitian realm \cite{el2018non,ozdemir2019parity,RevModPhys.93.015005,doi:10.1080/00018732.2021.1876991,OS23,wiersig2020review, parto2020non}. NH systems exhibit intriguing properties, being the most relevant to their complex spectrum with degeneracies known as exceptional points (EPs), where eigenvalues and eigenvectors coalesce
\cite{TKato, heiss2004exceptional, berry2004physics, Heiss_2012, PhysRevLett.86.787, PhysRevLett.103.134101, PhysRevLett.104.153601, gao2015observation, doppler2016dynamically,PhysRevB.99.121101}.  EPs have shown to be topological objects with topology defined in the complex spectrum, thus introducing the concept of point gaps unlike Hermitian topology \cite{SZF18, PhysRevX.9.041015,KBS19,ZSHC21}.  Also, the NH spectrum is very sensitive to perturbations, causing strong responses at EPs and offering a way for sensing devices with no Hermitian analog \cite{wiersig2020review,PhysRevLett.125.180403,arouca2022exceptionally,parto2023enhanced}.

NH physics has been shown to emerge in open systems \cite{RevModPhys.93.015005,doi:10.1080/00018732.2021.1876991}, with well-established mechanisms in photonics and optics where non-Hermiticity can be controlled almost at will \cite{PhysRevLett.86.787,feng2014single,gao2015observation,doppler2016dynamically,peng2016chiral,hodaei2017enhanced,chen2017exceptional,Longhi:17}. Another less explored but promising route is offered by material junctions where non-Hermiticity naturally arises when coupling a closed system to normal reservoirs \cite{datta1997electronic}. These ideas have recently been explored in normal systems \cite{PhysRevB.98.155430, PhysRevB.98.245130, PhysRevResearch.1.012003,cayao2023exceptional,NIOS23} and a few examples also exist in setups with superconductors \cite{pikulin2012topological,PhysRevB.87.235421,JorgeEPs,avila2019non,PhysRevB.105.094502,PhysRevB.107.104515,PhysRevB.105.155418,cayao2024nonhermitian}. While these initial studies have shed some light on the interplay between non-Hermiticity and superconductivity, they have overlooked the impact of non-Hermiticity on the phase of the superconducting order parameter. 

The most straightforward systems where the superconducting phase is crucial are Josephson junctions (JJs), formed by two superconductors with their order parameters exhibiting a finite phase difference \cite{RevModPhys.51.101,Tinkham}. This enables  the formation of Andreev bound states (ABSs) and the transfer of Cooper pairs, giving rise to a dissipationaless supercurrent across the JJ \cite{kulik1975,furusaki1991dc,PhysRevB.45.10563,Beenakker:92,Furusaki_1999,RevModPhys.76.411,sauls2018andreev}. These properties make JJs excellent devices for revealing the type of  superconductivity \cite{tiira17,PhysRevLett.121.047001,PhysRevX.9.011010,ren2019topological,Fornieri_2019,PhysRevLett.124.226801,PhysRevLett.125.116803,lutchyn2018majorana,prada2019andreev,frolov2019quest,flensberg2021engineered}  and also essential superconducting building blocks for future quantum  applications 
\cite{PhysRevLett.89.117901,devoret2004superconducting,PhysRevLett.90.226806,PhysRevB.81.144519,sarma2015majorana,acin2018quantum,krantz2019quantum,aguado2020perspective,benito2020hybrid,aguado2020majorana,beenakker2019search,PRXQuantum.2.040204,siddiqi2021engineering,bargerbos2022singlet,doi:10.1146/annurev-conmatphys-031119-050605,pita2023direct}. Thus, even though JJs offer a natural ground for NH physics, it is still unknown the response of ABSs and supercurrents to non-Hermiticity, specially, to the presence of EPs.  

\begin{figure}[!t]
\centering
	\includegraphics[width=0.49\textwidth]{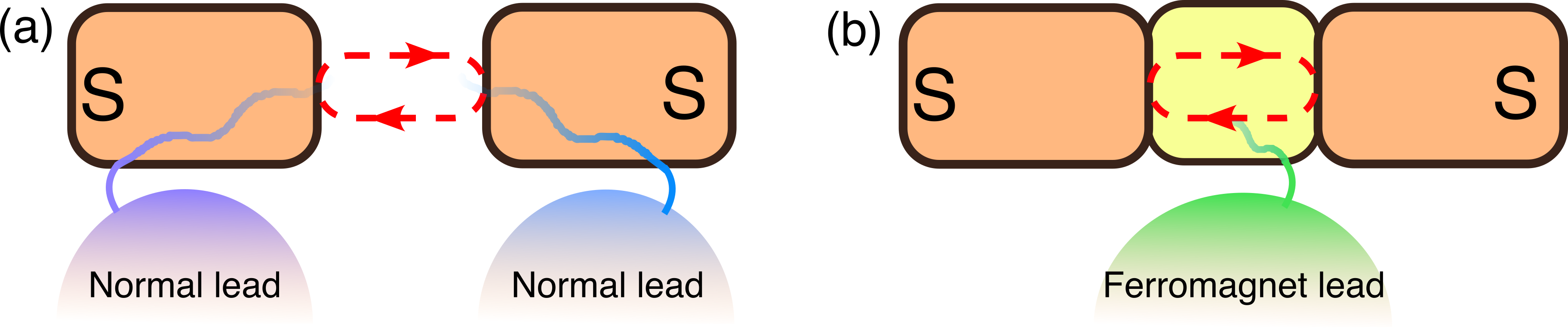}
 	\caption{(a) A NH JJ formed by coupling two superconductors (orange)  without any mediating region to different normal leads . (b) A NH JJ formed by coupling a very short mediating normal region (yellow)  to a ferromagnet lead. In both cases, the emerging ABSs (red dashed loop between superconductors) feel the presence of non-Hermiticity, indicated by shaded lines inside of the orange and yellow regions.}
\label{Fig1} 
\end{figure}

In this work, we consider NH phase-biased JJs formed by conventional superconductors coupled to normal leads [Fig.\,\ref{Fig1}] and discover topologically robust EPs and EP-enhanced supercurrents fully controlled by the phase difference. We find that the ABSs   reveal the formation of EPs at  finite and zero real energies. Interestingly, their phase-dependent supercurrents become sharply enhanced   at  EPs, but only the EPs at zero real energy produce sizeable signals in the total supercurrents. Our work thus demonstrates the potential of NH topology for realizing   EPs and enhanced supercurrents  in phase-biased JJs.

\textit{NH Josephson junctions}.---We consider JJs under non-Hermiticity induced by normal leads, as shown in Fig.\,\ref{Fig1}. We model these JJs by an effective Hamiltonian given by
\begin{equation}
\label{Heff}
H_{\rm eff}=H_{\rm JJ}+\Sigma^{r}(\omega=0)\,,
\end{equation}
where $H_{\rm JJ}$ describes the Hermitian JJ with superconductors (Ss) having a finite phase difference ($\phi$) between their order parameters \cite{PhysRevB.54.7366,zagoskin}. $\Sigma^{r}(\omega=0)$ is the retarded zero-frequency self-energy originating from the normal (ferromagnet) leads, which renders the JJ NH, as derived in Supplementary Material (SM) \cite{SM}. In the Hermitian regime,  the spectrum of $H_{\rm JJ}$ acquires a strong dependence on  $\phi$ and reveals the ABSs within the induced superconducting gap, producing a dissipationless supercurrent $I(\phi)$ \footnote{Hermitian JJs formed by conventional superconductors exhibit $I(\phi)\sim{\rm sin(\phi)}$ \cite{RevModPhys.51.101,RevModPhys.76.411} but more exotic ABSs can induce anomalous $I(\phi)$\cite{PhysRevB.53.R11957,PhysRevB.56.892,kashiwaya2000tunnelling,PhysRevLett.96.097007,PhysRevLett.103.107002,PhysRevB.92.155437,mizushima2018multifaceted,PhysRevB.101.184501,PhysRevB.104.L020501,PhysRevB.105.054504,baldo2023zero}. We also note that Ss forming JJs can be coupled directly or by a mediating region [Fig.\,\ref{Fig1}(a,b)]: when the length of the mediating region is shorter than the superconducting coherence length, only a pair of ABSs appears within the gap which then fully determines $I(\phi)$ \cite{Beenakker:92}.}.    
NH effects arise from  $\Sigma^{r}(\omega=0)$ \cite{datta1997electronic}, which  
in general has real (Re) and imaginary (Im) parts, $\Sigma^{r}(\omega=0)={\rm Re}\Sigma^{r}-i{\rm Im}\Sigma^{r}$ \cite{SM}.  While ${\rm Re}\Sigma^{r}$ only causes a shift in the energies of  $H_{\rm JJ}$,  ${\rm Im}\Sigma^{r}$ makes $H_{\rm eff}$  NH, inducing changes in the spectrum that are entirely due to the NH mechanism.

Before going further, we stress the relation between ABSs and $I(\phi)$. The spectrum, and hence ABSs, can be obtained from ${\rm det}(\omega-H_{\rm eff})=0$, which corresponds to the poles of the retarded Green's function \cite{datta1997electronic,mahan2013many,zagoskin} $G^{r}(\omega)=(\omega-H_{\rm eff})^{-1}$, where $H_{\rm eff}$  is given by Eq.\,(\ref{Heff}). Due to non-Hermiticity, the spectrum becomes complex with eigenvalues coming in pairs due to particle-hole symmetry  \cite{pikulin2012topological,PhysRevB.87.235421,JorgeEPs,avila2019non,PhysRevB.105.094502,PhysRevB.107.104515}, namely, ($E_{n},-E_{n}^{*}$), $E_{n} (\phi)=  {\rm Re}E_{n}(\phi)-i{\rm Im}E_{n}(\phi)$. The Re part represents the quasiparticle energy while the Im part characterizes its lifetime  \cite{datta1997electronic}. 
Since the Josephson current is an equilibrium flow in the junction, the equilibrium spectrum of $H_{\rm eff}$ determines the Josephson current.
Thus, in the same spirit as in  Hermitian JJs, where the  Andreev spectrum determines the supercurrent \cite{furusaki1991dc,PhysRevB.45.10563,Beenakker:92},  it is natural here to define a supercurrent associated with the complex spectrum of NH JJs as \cite{PhysRevX.8.031079} $I=-(e/\hbar)\sum_{n\ge0}dE_{n}(\phi)/d\phi$, which leads to
\begin{equation}
\label{INH}
I(\phi)=\frac{e}{\hbar}\sum_{n\le0}\left[\frac{d {\rm Re}E_{n}(\phi)}{d \phi}-i  \frac{d [{\rm Im}E_{n}(\phi)]}{d \phi}\right]\,,
\end{equation}
where we take the summation over $n$ for the occupied Andreev states with ${\rm Re}E_{n}\le0$ \footnote{It is common to find versions of Eq.\,(\ref{INH}) with a factor of $2e/\hbar$ instead of $\frac{e}{\hbar}$, which occurs because spin degeneracy is usually assumed, see e. g.\,Refs.\,\cite{Beenakker:92,PhysRevB.96.205425}. However, we consider effective spinless models   with broken spin-degeneracy.}. 
As shown in the SM \cite{SM}, the physical supercurrent is ${\rm Re}I(\phi)$.  Below, we consider two types of NH JJs with non-Hermiticity due to coupling to normal and ferromagnet leads and study EPs in the complex Andreev spectrum and their impact on the supercurrent $I(\phi)$.
 
\begin{figure}[!t]
\centering
	\includegraphics[width=0.49\textwidth]{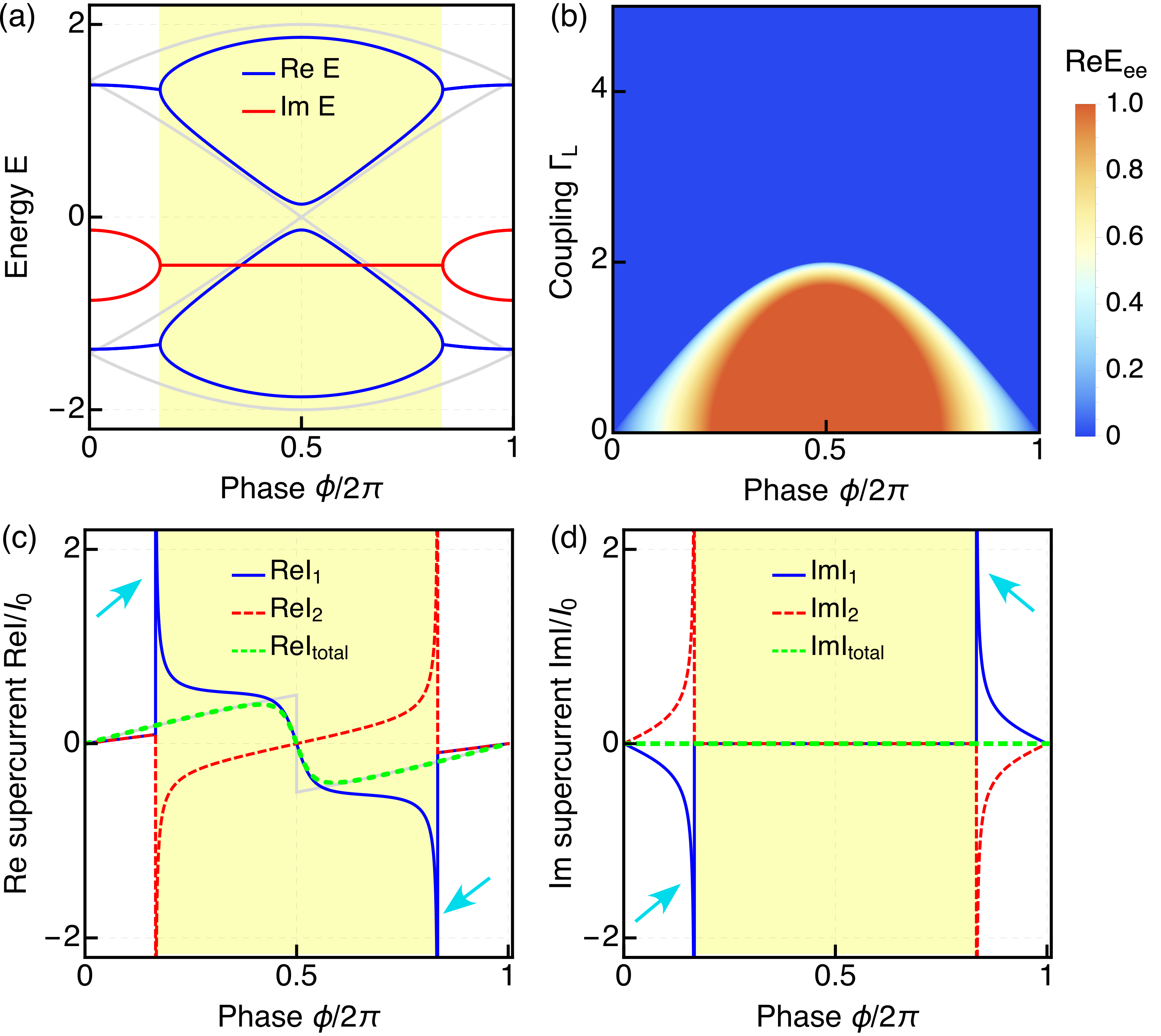}
 	\caption{(a) Re (blue) and Im (red) parts of the eigenvalues given by Eq.\,(\ref{Eval1}) as a function of  $\phi$. (b) Re part of the difference between the two positive eigenvalues as a function of $\Gamma_{\rm L}$ and $\phi$, with the blue region borders marking the  EPs. (c,d) Total (green) and individual (blue and red)  supercurrents $I(\phi)$ due to occupied ABSs as a function of $\phi$. Cyan arrows indicate the divergence of $I(\phi)$ at the EPs. Gray curves in   (a,c) correspond to eigenvalues and total $I(\phi)$ at $\Gamma_{L,R}=0$. Parameters: $\Delta=1$, $t=1$, $\varepsilon_{\alpha}=0$, $\Gamma_{L}=1$, $\Gamma_{R}=0$, $I_{0}=e\Delta/\hbar$.}
\label{Fig2} 
\end{figure}

\textit{NH JJ with superconductors coupled to different normal leads}.---To begin, we consider the NH JJ with single site superconductors (Ss) coupled to distinct normal (N) leads, as shown in Fig.\,\ref{Fig1}.
Equation (\ref{Heff}) 
with
 \begin{equation}
 \label{NHJJtoy1}
H_{\rm JJ}^{(1)}=
\begin{pmatrix}
H_{L}&V\\
V^{\dagger}&H_{R}
\end{pmatrix}\,,\quad
\Sigma^{(1)}={\rm diag}(\Sigma_{L},\Sigma_{R})\,,
\end{equation} 
describes ABSs in this NH JJ,  where $H_{\alpha}=\varepsilon_{\alpha}\tau_{z}+{\rm Re}(\Delta_{\alpha})\tau_{x}-{\rm Im}(\Delta_{\alpha})\tau_{y}$ for $\alpha=L$ $(R)$ is the Bogoliubov de-Gennes Hamiltonian of the left (right) S with an $s$-wave pair potential $\Delta_{\alpha}=\Delta{\rm e}^{i\phi_{\alpha}}$ and the onsite normal energy $\varepsilon_\alpha$, while $V=t\tau_{z}$ is the coupling between Ss  \footnote{The parameter $t$ controls the transmission across the junction, enabling to reach vanishing transmission and transmission equal to 1 which, respectively, characterize the tunnel and full transparent regimes \cite{Beenakker:92,PhysRevB.54.7366,datta1997electronic}.}. The self-energy term gives non-Hermiticity $\Sigma_{L(R)}=-i\Gamma_{L(R)}\tau_{0}$  \footnote{Different couplings $\Gamma_{L(R)}$ give a finite asymmetry in the amount of non-Hermiticity, shown before to be the key for inducing EPs \cite{PhysRevB.105.094502,PhysRevB.107.104515}}, where 
the coupling of the left (right) S to the left (right) N lead determines the decay rate $\Gamma_{L(R)}$  \cite{PhysRevB.105.094502,PhysRevB.107.104515,PhysRevB.105.155418}. ($\tau_{i}$  is the $i$-th Pauli matrix in the Nambu space.) For $\varepsilon_{\alpha}={0}$, $\phi_{L}=0$, and $\phi_{R}=\phi$, the eigenvalues of $H_{\rm eff}$ read   
\begin{equation}
\label{Eval1}
E_{j}=-i\Gamma\pm\sqrt{\delta^{2}-\gamma^{2}\pm2\Delta\sqrt{\smash[b]{t^{2}{\rm sin}^{2}(\phi/2)-\gamma^{2}}}}
\end{equation}
where $\delta^{2}=t^{2}+\Delta^{2}$, 
$\gamma=(\Gamma_{L}-\Gamma_{R})/2$, $\Gamma=(\Gamma_{L}+\Gamma_{R})/2$, and $j=1,\cdots,4$ labels the eigenvalues.  Fig.\,\ref{Fig2}(a) shows the Re and Im parts of $E_j$ as a function of $\phi$ without and with non-Hermiticity. 
For $\Delta=t$, $E_j$ without non-Hermiticity shows level crossings at $\phi=0, \pi$, which qualitatively reproduces the standard ABSs in transparent JJs  \cite{SM}. 
Notably, once turning on $\gamma\neq0$, the modes crossing at $\phi=0$ coalesce, with their Re parts merging at finite energy, signaling the formation of EPs \cite{RevModPhys.93.015005}. See yellow region ends in  Fig.\,\ref{Fig2}(a). From  Eq.\,(\ref{Eval1}), EPs appear when $t^{2}{\rm sin}^{2}(\phi/2)=\gamma^{2}$, and thus, they exist even for small but nonzero $\gamma$.
One can also confirm this feature in Fig.\,\ref{Fig2}(b), which shows the Re part of the difference between the two positive eigenvalues (${\rm Re}E_{\rm ee}$)  as a function of $\gamma$ and $\phi$. 
The border of the blue region (${\rm Re}E_{\rm ee}=0$) marks the EPs, 
which persist for $|\gamma|<|t|$. In contrast, the modes crossing at $\phi=\pi$ do not evolve into EPs.  The system has parity-time, spin-rotation, and particle-hole symmetries. An NH topology protected by these symmetries ensures the presence and absence of these EPs \cite{SM}; thus, the discussed EP feature is universal.

 We next analyze the current-phase curve $I(\phi)$ for the ABSs in Eq.\,(\ref{Eval1}), obtained from Eq.\,(\ref{INH}). See Fig.\,\ref{Fig2}(c,d).   In the Hermitian regime, $I(\phi)$  exhibits a sawtooth profile at $\phi=\pi$ due to the zero-energy crossing in the ABSs, as the gray curve indicates. In the NH case, with $\gamma\neq0$, the ${\rm Re}I(\phi)$  and ${\rm Im}I(\phi)$ of each ABS develop sharp divergences at the EP since each ABS carries $\smash[b]{I_{1(2)}(\phi)}\sim \pm {\rm sin}(\phi)/\sqrt{\smash[b]{t^{2}{\rm sin}^{2}(\phi/2)-\gamma^{2}}}$, whose denominator vanishes at the EPs \cite{SM}. See red and blue curves in (a,b) and cyan arrows. 
 However, the divergences of $I_1(\phi)$ and $I_2(\phi)$ cancel out in the total  ${\rm Re}I(\phi)$, leading to a smooth curve across EPs with an overall sine-like profile, as shown by the green dotted curve  \cite{SM}; similarly, the total ${\rm Im}I(\phi)$  acquires vanishing values [Fig.\,\ref{Fig2}(d)]. Nevertheless, it is worth noting that, even though no EP effect is evident in the total $I(\phi)$, the Hermitian sawtooth profile at $\phi=\pi$  smooths out by $\gamma\neq0$.  In summary,  NH JJs can host tunable EPs at finite Re energies as a pure NH topological effect, but the total supercurrent is insensitive to such EPs.

\begin{figure}[!t]
\centering
\includegraphics[width=0.49\textwidth]{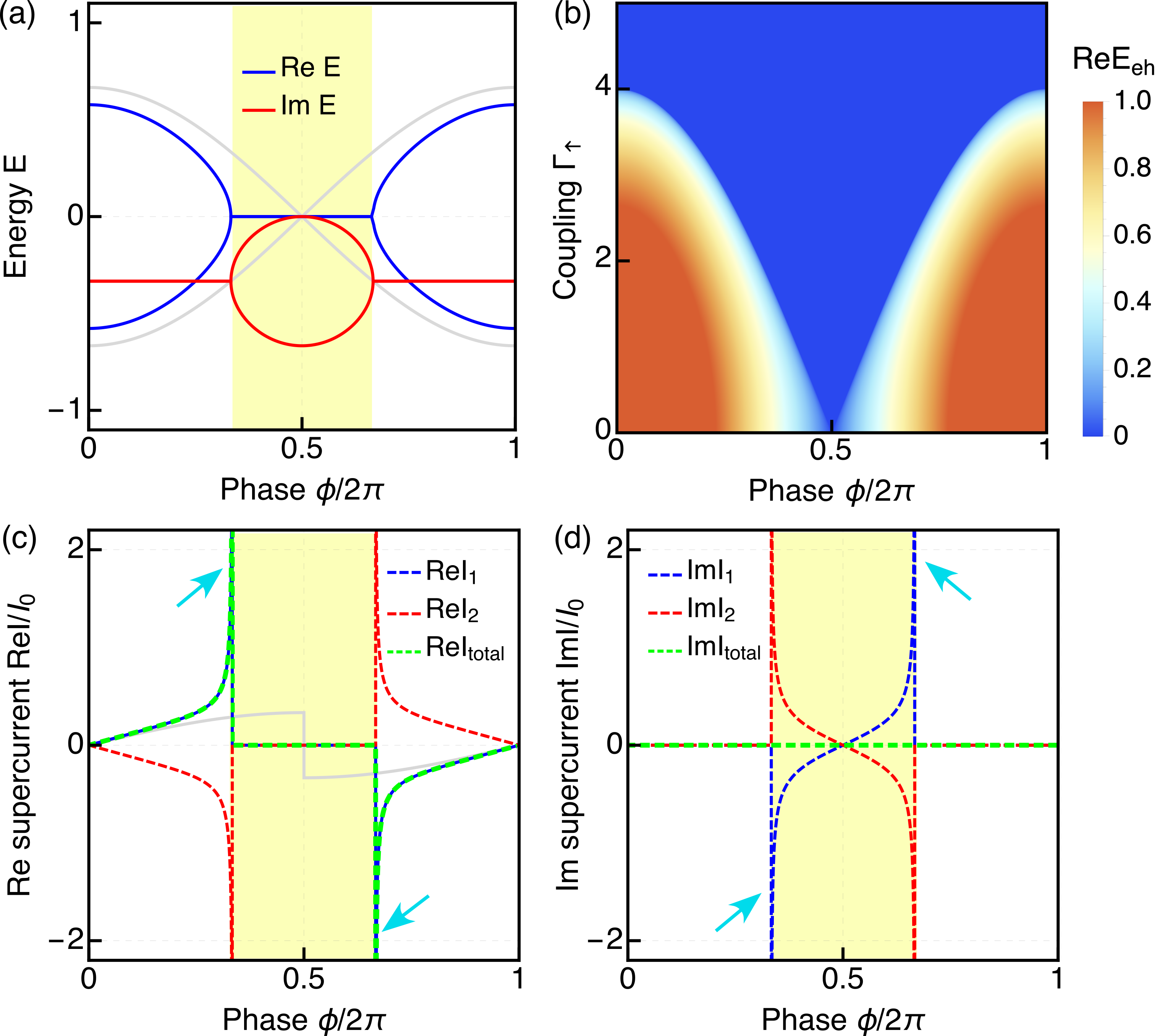}
\caption{(a) Re (blue) and Im (red) parts of the eigenvalues given by Eq.\,(\ref{Eval2}) as a function of  $\phi$. (b) Re part of the difference between the two lowest positive and lowest negative eigenvalues as a function of $\Gamma_{\uparrow}$ and $\phi$, with the blue region borders marking the  EPs.  (c,d) Total (green) and individual (blue and red)   $I(\phi)$ due to both ABSs as a function of $\phi$. Cyan arrows show the enhancement of $I(\phi)$ at the EPs.  Gray curves in  (a,c) correspond to eigenvalues and   $I(\phi)$ at $\Gamma_{\uparrow,\downarrow}=0$. Parameters: $\Delta=1$, $t=1$, $\varepsilon_{\alpha}=0$, $\Gamma_{\uparrow}=2$, $\Gamma_{\downarrow}=0$, $I_{0}=e\Delta/\hbar$.}
\label{Fig3} 
\end{figure}

 \textit{NH JJ with a middle N region coupled to a ferromagnet lead}.---Here we consider spinful NH JJs with a short mediating N region coupled to a ferromagnet lead, as shown in Fig.\,\ref{Fig1}(b). For simplicity, we focus on single site S and N regions  \footnote{JJs with a very small N region belongs to the short junction regime where the length of the N region is smaller than the superconducting coherence length \cite{Beenakker:92}. The same applies to the JJ studied in Eq.\,(\ref{NHJJtoy1})}. The NH  JJ   is then modeled by
\begin{equation}
\label{NH3}
H_{\rm JJ}^{(2)}=
\begin{pmatrix}
H_{L}&V&0\\
V&H_{N}&V\\
0&V&H_{R}
\end{pmatrix}\,,\quad
\Sigma^{(2)}={\rm diag}(0,\Sigma,0)\,,
\end{equation}
where $H_{\alpha}=\varepsilon_{\alpha}\tau_{z}+{\rm Re}(\Delta_{\alpha})\sigma_{y}\tau_{y}-{\rm Im}(\Delta_{\alpha})\sigma_{y}\tau_{x}$ for $\alpha=L$ $(R)$ describes the left (right) S with an $s$-wave pair potential $\Delta_{\alpha}=\Delta{\rm e}^{i\phi_{\alpha}}$, $H_{\rm N}=\varepsilon_{\rm N}\tau_{z}$ presents the N region, and $V=t\sigma_{0}\tau_{z}$ is the hopping between N and S  \footnote{Here, $t$ controls the transmission across the junction \cite{Beenakker:92,PhysRevB.54.7366,datta1997electronic}.}. 
($\sigma_i$ and $\tau_i$ are the Pauli matrices in the spin and Nambu spaces.)
The coupling between N and a ferromagnet lead induces a spin-dependent self-energy $\Sigma={\rm diag}(\Sigma^{r}_{\rm e},\Sigma^{r}_{\rm h})$ where $\Sigma^{r}_{\rm e,h}(\omega=0)=-i \Gamma \sigma_{0}-i\gamma \sigma_{z}$, $\Gamma=(\Gamma_{\uparrow}+\Gamma_{\downarrow})/2$, $\gamma=(\Gamma_{\uparrow}-\Gamma_{\downarrow})/2$;   $\Gamma_{\sigma}$ is the coupling of spin $\sigma$   to the lead \footnote{The choice of non-Hermiticity here is different in comparison to the previous section, where the asymmetry therein originated from coupling  Ss to different N leads.}.  The eigenvalues of $H_{\rm eff}$ are not simple, but it is still possible to explore them as solutions of ${\rm det}(\omega-H_{\rm JJ}^{(2)}-\Sigma^{(2)})=0$. Taking up to second order in $\omega$, the solutions for $\varepsilon_{\alpha}=0$, $\phi_{L}=0$, $\phi_{R}=\phi$ read
\begin{equation}
\label{Eval2}
E_{\pm}=-i\Gamma \frac{\Delta^{2}(2t^{2}+\Delta^{2})}{A}\pm\frac{\Delta}{A}\sqrt{\smash[b]{C+2t^{4} A\,{\rm cos}\phi}}\,,
\end{equation}
where $A=4 t^4+\Delta ^4+2 \Delta ^2 \left(-\gamma ^2+\Gamma ^2+2 t^2\right)$, and 
$C=2t^4 A-4t^2\gamma^2\Delta^4+2(\gamma^2+\Gamma^2)^2\Delta^4-\gamma^2\Delta^6$.
From Eq.\,(\ref{Eval2}), we find that the ABSs are degenerate when $C+2t^{4} A\,{\rm cos}\phi=0$.
In the absence of non-Hermiticity ($\Gamma=\gamma=0$), it holds that $C=2t^4A>0$; thus, the degeneracy appears at $\phi=\pi$ as a level crossing at zero energy, as illustrated by gray curves in Fi.\ref{Fig3} (a). 
Then, in the presence of non-Hermiticity, the degeneracy evolves into a pair of EPs at $\phi_{\rm EP}=\pm {\rm arccos}(-C/2t^4 A)$ once we have $|C|<2t^4|A|$ by non-Hermiticity \footnote{We have verified that the corresponding wavefunctions also coalesce at these EPs.}.  
Intriguingly, the Re part of ABSs sticks to zero energy between the EPs, unlike the EPs in the previous model.
We note that a weak but nonzero spin-dependent non-Hermiticity $\gamma$ suffices to realize the EPs because $C$ decreases as $C=2t^2A-4t^2\gamma^2\Delta^4$ for $0< |\gamma|, |\Gamma|\ll |t|, |\Delta|$ so to meet the condition $|C|<2t^4|A|$.
The tunability of the EPs is also evident in Fig.\,\ref{Fig3}(b), which shows the Re part of the difference between the eigenvalues $E_{\pm}$ as a function of $\Gamma_{\uparrow}$ and $\phi$; EPs appear at weak $\Gamma_{\rm \uparrow}$. The border of the blue region marks the EPs. 
Contrary to the previous system, this JJ breaks parity-time and spin-rotation symmetries and only keeps particle-hole symmetry intrinsic to superconductors. 
This difference enables the topologically stable EPs at zero energy,
as argued in the SM \cite{SM}.

When it comes to the supercurrent   $I(\phi)$  due to ABSs in Eq.\,(\ref{Eval2}), Fig.\,\ref{Fig3}(c,d)   shows its Re and Im parts as a function of $\phi$. In the Hermitian case,  $I(\phi)$ develops a sawtooth profile at $\phi=\pi$ due to the zero-energy crossing in the ABSs; see the gray curve. In the NH regime, the Re and Im parts of $I(\phi)$ interestingly develop divergences at the EPs of the ABSs; see the cyan arrows.  Here, $I(\phi)\sim {\rm sin}(\phi)/\sqrt{\smash[b]{C+2t^{4} A\,{\rm cos}\phi}}$, whose denominator thus vanishes at the EPs \cite{SM}. At first sight, this seems similar to $I(\phi)$ from  EPs at finite Re energy in Fig.\,\ref{Fig2}(c,d), but there is a key difference. In the present case, ${\rm Re}I(\phi)$ is entirely determined by only the non-negative  ABS, which then implies that the enhancement at the EPs could be robust in contrast to $I(\phi)$ from EPs at finite Re energy  \footnote{While reaching the EPs in real systems might be challenging,  but being close to them
already induces a drastic enhanced $I(\phi)$.}. Moreover,  while the total ${\rm Im}I(\phi)$ vanishes because the two ABSs merging into a single level at zero Re energy (yellow region) necessarily involves taking both Im ABSs, the ${\rm Re}I(\phi)$ intriguingly vanishes between EPs as a result of the zero-energy lines in the Re ABSs. The fact that $I(\phi)$  has robust enhancement at EPs fully controlled by $\phi$ reveals that this effect has a  pure  NH origin.

\textit{NH JJs in superconductor-semiconductor hybrids}.---Having found  EPs and EP-enhanced $I(\phi)$ in simple NH JJs, here we show their realization in experimentally feasible NH Rashba JJs. The Hermitian part is modelled by $H_{\rm R}=\xi_{k}\tau_{z}+i\alpha k \sigma_{y}\tau_{z}+B\sigma_{x}\tau_{z}+\Delta\sigma_{y}\tau_{y}$, where $\xi_{k}=(\hbar^{2}k^{2}/2m-\mu)$ is the kinetic term with wave number $k$ along $x$, $\mu$ is the chemical potential, $\alpha$ is the Rashba spin-orbit coupling (SOC), and $\Delta$  is the  spin-singlet $s$-wave pair potential.   Since SOC is intrinsic  in  semiconductors, we consider it to be finite \footnote{Similar results can be obtained without SOC but its intrinsic nature in semiconductors makes it    necessary in realistic systems \cite{sato2017topological,Aguadoreview17,prada2019andreev,lutchyn2018majorana,frolov2019quest,flensberg2021engineered,tanaka2024theory}.}. $B$ is a Zeeman field   along   $x$  inducing a Hermitian topological phase transition at $|B_c|=\sqrt{\smash[b]{\Delta^{2}+\mu^{2}}}$ that predicts Majorana bound states (MBSs) \cite{PhysRevLett.105.077001,PhysRevLett.105.177002,sato2017topological,Aguadoreview17,prada2019andreev,lutchyn2018majorana,frolov2019quest,flensberg2021engineered,tanaka2024theory}. Then, $H_{\rm R}$ is discretized into  a tight-binding lattice with spacing $a=10$\,nm  \cite{cayao2018andreev} and   divided into three finite regions: The left and right S regions host a  finite pair potential $\Delta$ with a finite phase difference $\phi$, while the central N region has $\Delta=0$; each region has length $L_{\rm N, S}$ and  the chemical potential $\mu_{\rm N, S}$. We take realistic parameters: $\alpha_{\rm R}=40$\,meVnm and $\Delta=0.5$\,meV, in the range of values for InSb and InAs nanowires, and Nb and Al Ss \cite{lutchyn2018majorana}. We finally model a short NH JJ by coupling the middle N region to a ferromagnet lead as in Eq.\,(\ref{NH3}) [Fig.\,\ref{Fig1}(b)], and obtain the Andreev spectrum and $I(\phi)$.
  
\begin{figure}[!t]
\centering
	\includegraphics[width=0.49\textwidth]{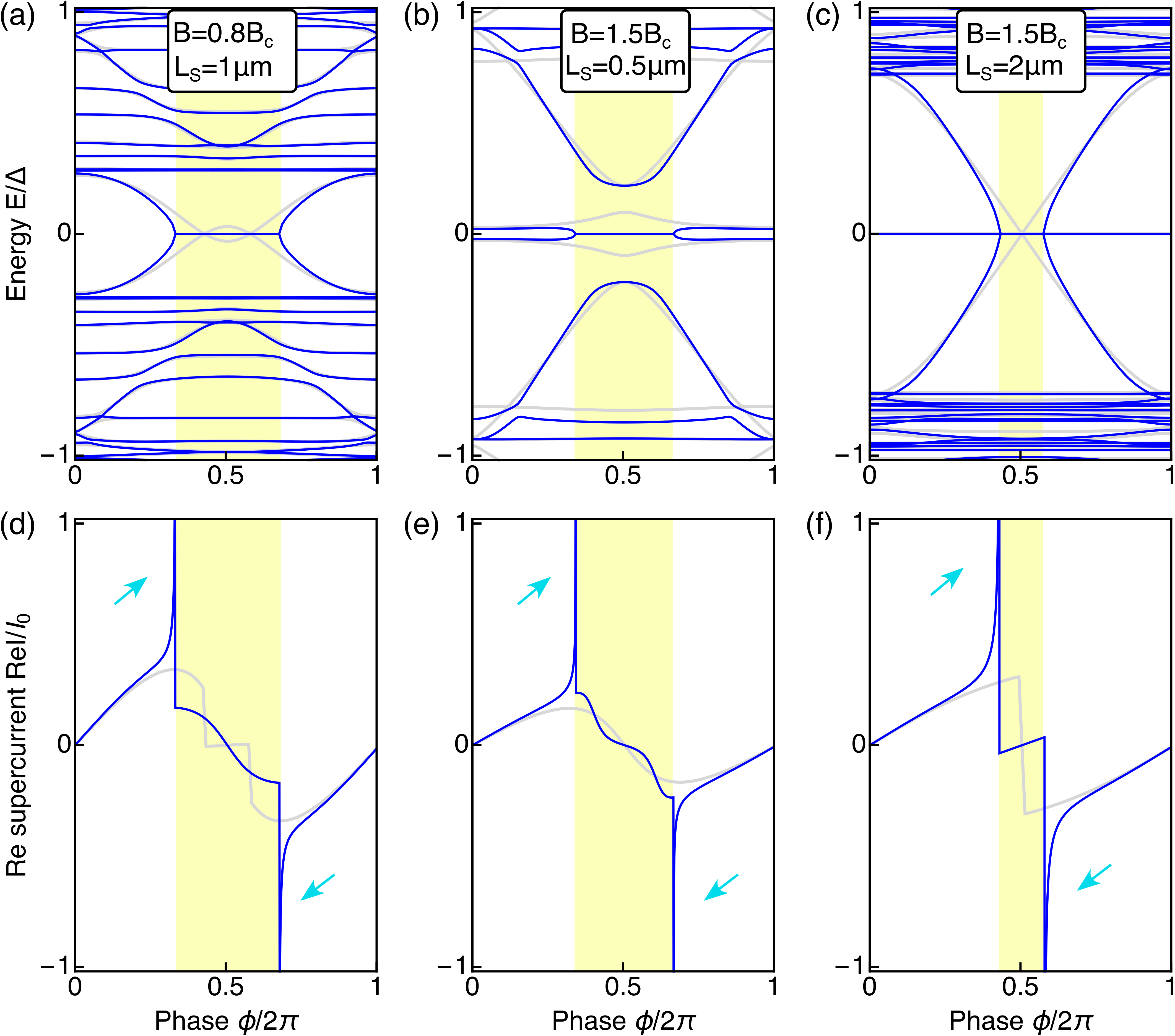}
 	\caption{(a-c) Re  spectrum (blue) of a   short  NH Rashba JJ  as a function of  $\phi$ for different $B$ and $L_{\rm S}$. In (a) $\Gamma_{\uparrow}=1.5$meV, while for (b,c)  $\Gamma_{\uparrow}=3.5$meV. Gray curves correspond to  $\Gamma_{\uparrow,\downarrow}=0$. The yellow region ends   mark the   EPs   at zero energy. (d-f) ${\rm Re}I(\phi)$ for  (a-c) as a function of $\phi$ with (blue) and without (gray) non-Hermiticity. Cyan arrows indicate the  ${\rm Re}I(\phi)$ at EPs. Parameters: $\alpha=40$meVnm, $\Delta=0.5$meV, $\mu_{\rm N(S)}=0.5$meV, $L_{\rm N}=20$nm, $\Gamma_{\downarrow}=0$, $I_{0}=e\Delta/\hbar$.}
\label{Fig4} 
\end{figure}

In Fig.\,\ref{Fig4}, we present the Re Andreev spectrum and ${\rm Re}I(\phi)$ as a function of $\phi$. For $B<B_{\rm c}$ in the Hermitian regime, the JJ develops zero-energy crossings, which evolve into stable zero-energy lines at finite spin-dependent non-Hermiticity $\gamma\neq0$ whose ends mark the presence of EPs. See the gray and blue curves and also the yellow region in Fig.\,\ref{Fig4}(a). These EPs do not depend on the length of the S regions
\footnote{MBSs appear as end states, with their wavefunctions decaying towards the bulk and overlapping in short systems \cite{PhysRevLett.105.077001,PhysRevLett.105.177002,sato2017topological,Aguadoreview17,prada2019andreev,lutchyn2018majorana,frolov2019quest,flensberg2021engineered}. This finite overlap is also accompanied with a zero-energy splitting, which becomes vanishing small for long systems and is thus seen as a signature of Majorana nonlocality \cite{PhysRevB.96.205425,PhysRevB.104.L020501}.}.  Above $B_{\rm c}$, the Hermitian JJ hosts four (nearly zero energy) MBSs at $\phi=\pi$ (two at the inner side of the junction and two at the outer ends) but only two at $\phi=0$ (two at the outer), as revealed in the gray curves in (b) or (c) \cite{PhysRevLett.108.257001,PhysRevB.91.024514,cayao2018andreev,cayao2018finite,PhysRevB.96.205425}. Interestingly, for the short S case $L_{\rm S}<2\xi_{\rm M}$, with $\xi_{\rm M}$ the Majorana localization length, a finite amount of non-Hermiticity induces EPs between lowest positive and highest negative ABSs (outer MBSs) and form a zero-energy line between them [Fig.\,\ref{Fig4}(b)]. For $L_{\rm S}>2\xi_{\rm M}$, the outer MBSs acquire zero-energy for all $\phi$ and the inner MBSs develop a zero-energy crossing at $\phi=\pi$ in the Hermitian regime (gray curves), which, at $\gamma\neq0$, transforms into a pair of EPs between inner MBSs connected by a stable zero-energy line  [Fig.\,\ref{Fig4}(c)]. This EP regime is analogous to the case found in  Fig.\,\ref{Fig3}(a).   

In Fig.\,\ref{Fig4}(e-f), we show $I(\phi)$ of Eq.\,(\ref{INH}) for the Andreev spectrum in Fig.\,\ref{Fig4}(a-c), which includes the contribution from ABSs and the quasicontinuum. In  all the cases, we find divergent profiles at EPs, which remain robust even under the presence of a phase-dependent quasicontinuum, as indicated by cyan arrows in Fig.\,\ref{Fig4}(d-f). Note that EPs between inner MBSs produce larger supercurrents for $L_{\rm S}>2\xi_{\rm M}$ since the quasicontinuum is weakly dependent on $\phi$ so the MBSs carry almost all the contribution. As argued above, this surprising behavior,  the enhancement of Re $I(\phi)$ at EPs,  originates from the anomalous contributions of ABSs at EPs occurring at zero-real energy and represents a unique and robust NH effect \footnote{Here EPs at finite energy might emerge accidentally, but their respective supercurrents do not sense the EPs, in line with the first simple model.}. 

 \textit{Impact of the imaginary energy on the supercurrent}.---We have shown that the intrinsic contribution of EPs to the Josephson current comes from the Re energy and develops a divergent profile. While this aspect reveals the intrinsic EP contribution, the Im part of the energy results in the band broadening and obscures each energy band’s distinction. In particular, the band broadening mixes an occupied state with a negative real energy and its particle-hole partner empty state with a positive real energy near the exceptional point. Since an occupied state and its particle-hole partner contribute to the Josephson current with an opposite sign, the mixing suppresses the divergence in the Josephson current. Still, we can show that the exceptional point enhances the Josephson current. To show the enhancement, let us consider a typical occupied Andreev state near an EP
\begin{align}
E_n(\phi)=-\sqrt{s(\phi)}-i\Gamma\,,    
\end{align}
where $s(\phi)$ is given by
\begin{align}
s(\phi)=a^2\cos^2(\phi/2)-b\gamma^2, \quad (a>0, \, b\ge 0)\,.
\label{eq:s}
\end{align}
Here $a$ has the same dimension as $\gamma$ and $b\ge 0$ is dimensionless.
At $\phi=\phi_0$ with $s(\phi_0)=0$,
$E_n(\phi)$ degenerates with the particle-hole partner $-E_n^*(\phi)$.
For $b=0$, the degeneracy is a simple level crossing, while for $b>0$, the degeneracy is an EP with a branch cut of square root.
Because the mixing becomes significant when $|{\rm Re}E_n(\phi)|$ is comparable with the imaginary part $|{\rm Im}E_n(\phi)|=\Gamma$, the current growth near the exceptional point terminates at $\phi_1$ satisfying $\sqrt{s(\phi_1)}=\Gamma$.
Thus, for $\gamma, \Gamma\ll a$, we can estimate the maximal current as
\begin{align}
I(\phi)|_{\rm max}\sim \frac{e}{\hbar}\left.\frac{d{\rm Re}E_n(\phi)}{d\phi}\right|_{\phi=\phi_1}=\frac{e}{\hbar}\frac{a}{2}
\sqrt{1+b(\gamma/\Gamma)^2}\,,
\label{eq:max}
\end{align}
where  $\phi_1=\pi-(2\Gamma/a)\sqrt{1+b(\gamma/\Gamma)^2}$ \footnote{Whereas the formula in Ref.\cite{shen2024nonhermitian} gives a more accurate estimation of $I(\phi)|_{\rm max}$, the conclusion below is the same. See Sec.~S6 of \cite{SM}.}. 
We compare this estimation with the current without the EP. 
Since we can obtain the latter current by putting $b=0$ in Eq.(\ref{eq:max}), 
we conclude that the exceptional point enhances the Josephson current by a factor of $\sqrt{1+b(\gamma/\Gamma)^2}$.  We confirm this enhancement numerically in Sec.~S7 of Ref.\,\cite{SM}.

In conclusion, we studied non-Hermitian phase-biased Josephson junctions and demonstrated the formation of exceptional points at finite and zero real energies in the phase-dependent Andreev spectrum. We discovered that the Andreev exceptional points at zero real energy give rise to drastically enhanced phase-dependent supercurrents, thus boosting the sensing capability of Josephson junctions.  Our findings hold experimental relevance because similar non-Hermitian junctions have already been fabricated, including  Josephson junctions based on superconductor-semiconductor hybrids \cite{Doh:S05,nphys1811,Nilsson:NL12,PhysRevLett.110.217005,tiira17,Goffman17,PhysRevLett.121.047001,PhysRevX.9.011010,PhysRevLett.124.226801,PhysRevLett.125.116803,razmadze2022supercurrent} and few  Kitaev chains \cite{dvir2023realization,bordin2023crossed},  where coupling to normal leads seems to be a feasible task. Hence, these recent works   place our findings within experimental reach.  Our work thus opens a route for realizing exceptional points and enhanced supercurrents entirely by the interplay of non-Hermitian topology and the Josephson effect. 

\emph{Note added:} In the final stages of preparing this work, the preprint Ref.\,\cite{li2023anomalous}  was posted online,   partially overlapping with our results but using a different method and distinct setups.   We also note that a recent calculation by Shen et al. shows a similar enhancement of the Josephson currents by EPs \cite{shen2024nonhermitian}, consistent with our results  as we demonstrate in Sections 5, 6 and 7 of Ref.\,\cite{SM}.

We thank  R. Aguado and Y. Tanaka for their insightful discussions.   J. C. acknowledges financial support from the Swedish Research Council (Vetenskapsr{\aa}det Grant No. 2021-04121), the Royal Swedish Academy of Sciences (Grant No. PH2022-0003),  the Carl Trygger's Foundation (Grant No. 22: 2093), and the Japan Society for the Promotion of Science via the International Research Fellow Program. M. S. was supported by JST CREST Grant No. JPMJCR19T2.  The computations were enabled by resources provided by the National Academic Infrastructure for Supercomputing in Sweden (NAISS), partially funded by the Swedish Research Council through Grant Agreement No. 2022-06725.

\bibliography{biblio}
 
\onecolumngrid


\section*{Supplementary material}

# Supplemental Material for "Non-Hermitian phase-biased Josephson junctions"

Jorge Cayao[1] and Masatoshi Sato[2]

[1]*Department of Physics and Astronomy, Uppsala University, Box 516, S-751 20 Uppsala, Sweden*
[2]*Yukawa Institute for Theoretical Physics, Kyoto University, Kyoto 606-8502, Japan*
(Dated: June 27, 2024)

In this supplemental material we discuss the derivation of the effective Hamiltonian used in the main text to study non-Hermitian Josephson junctions and provide further analysis of their symmetries and non-Hermitian topology. Furthermore, we also discuss the Andreev bound states found in our models in the Hermitian regime and give analytical expressions for the supercurrents of the effective non-Hermitian models.

## S1. DERIVATION OF THE EFFECTIVE HAMILTONIAN

In the main text, we describe Josephson junctions (JJs) coupled to leads by an effective non-Hermitian Hamiltonian $H_{\rm eff}$. In this section, we derive the effective Hamiltonian in an exact manner. For this purpose, we consider a JJ and leads described by the Hamiltonians $H_{\rm JJ}$ and $H_{\rm L}$, respectively. The total Hamiltonian is

$$H = \begin{pmatrix} H_{\rm L} & T \\ T^\dagger & H_{\rm JJ} \end{pmatrix}, \tag{S1}$$

where $T$ is the hopping matrix between the JJ and the lead L. We assume that the lead is semi-infinite so we can treat it as an environment. From the retarded Green's function of the total system, we below derive the Green's function of the JJ under the influence of the lead, which gives the effective Hamiltonian.

From the equation of motion, we can show that the total retarded Green's function obeys the equation

$$(\omega - H)G^r(\omega) = 1, \tag{S2}$$

where we have omitted the infinite small imaginary part of $\omega + i\eta$ for simplicity. This equation is equivalent to

$$\begin{pmatrix} \omega - H_{\rm L} & -T \\ -T^\dagger & \omega - H_{\rm JJ} \end{pmatrix} \begin{pmatrix} G^r_{\rm L} & G^r_{\rm LJJ} \\ G^r_{\rm JJL} & G^r_{\rm JJ} \end{pmatrix} = \begin{pmatrix} 1 & 0 \\ 0 & 1 \end{pmatrix}, \tag{S3}$$

with $G^r_{\rm L(JJ)}$ the Green's function of the lead (JJ) and $G^r_{\rm LJJ/JJL}$ the Green's function between the lead and the JJ. In terms of the components, we have

$$\begin{aligned}
(\omega - H_{\rm L})G^r_{\rm L} - TG^r_{\rm JJL} &= 1, \\
(\omega - H_{\rm L})G^r_{\rm LJJ} - TG^r_{\rm JJ} &= 0, \\
-T^\dagger G^r_{\rm L} + (\omega - H_{\rm JJ})G^r_{\rm JJL} &= 0, \\
-T^\dagger G^r_{\rm LJJ} + (\omega - H_{\rm JJ})G^r_{\rm JJ} &= 1.
\end{aligned} \tag{S4}$$

Then, the second equation leads to

$$G^r_{\rm LJJ} = (\omega - H_{\rm L})^{-1}TG^r_{\rm JJ} = g_{\rm L}(\omega)TG^r_{\rm JJ}, \tag{S5}$$

where

$$g^r_{\rm L}(\omega) = (\omega - H_{\rm L})^{-1} \tag{S6}$$

is the Green's function of the isolated semi-infinite lead. Substituting Eq. (S5) into the last expression of Eqs. (S4), we obtain the Green's function of the JJ under the influence of the semi-infinite lead,

$$G^r_{\rm JJ} = [(\omega - H_{\rm JJ}) - T^\dagger g_{\rm L}(\omega)T]^{-1}, \tag{S7}$$

which defines the effective Hamiltonian of the JJ as

$$G^r_{\rm JJ} = [\omega - H_{\rm eff}]^{-1}. \tag{S8}$$

Thus, the effective Hamiltonian reads

$$H_{\text{eff}} = H_{\text{JJ}} + \Sigma^r(\omega), \tag{S9}$$

with the self-energy

$$\Sigma^r(\omega) = T^\dagger g_L^r(\omega) T. \tag{S10}$$

Therefore, the effective Hamiltonian given by Eq. (S9) incorporates the effects of the lead in an exact manner.

### S1.1. Self-energy

Having established that the incorporation of the leads is exact, we now discuss how to evaluate the self-energy. The main text discusses two types of systems: a system with coupled normal leads and a system coupled to a ferromagnet lead. Without loss of generality, we derive below the self-energy by the semi-infinite ferromagnet lead; a similar analysis leads to the self-energy by a semi-infinite normal lead.

As explained in the previous section, the evaluation of the self-energy requires the retarded Green's function of the isolated semi-infinite lead $g_L^r(\omega)$. Using the recursive Green's function approach, one can calculate the isolated semi-infinite lead's Green's function at the boundary. The Green's function at the boundary is diagonal in the Nambu space, with the electron (e) and hole (h) parts given by [1, 2]

$$g_{\sigma\sigma}^{e(h)}(\omega) = \begin{cases} \dfrac{1}{|t_L|} \left[ \dfrac{\omega - \epsilon_\sigma^{e(h)}}{2|t_L|} - \text{sgn}(\omega - \epsilon_\sigma^{e(h)}) \sqrt{\left(\dfrac{\omega - \epsilon_\sigma^{e(h)}}{2|t_L|}\right)^2 - 1} \right], & |(\omega - \epsilon_\sigma^{e(h)})/2|t_L|| > 1 \\ \dfrac{1}{|t_L|} \left[ \dfrac{\omega - \epsilon_\sigma^{e(h)}}{2|t_L|} - i\sqrt{1 - \left(\dfrac{\omega - \epsilon_\sigma^{e(h)}}{2|t_L|}\right)^2} \right], & |(\omega - \epsilon_\sigma^{e(h)})/2|t_L|| < 1, \end{cases} \tag{S11}$$

with $\sigma = \uparrow, \downarrow$, $\epsilon_\uparrow^e = -\epsilon_\uparrow^h = 2t_L - \mu_L + B$, and $\epsilon_\downarrow^e = -\epsilon_\downarrow^h = 2t_L - \mu_L - B$, where $t_L$ and $\mu_L$ are the next nearest neighbor hopping and the chemical potential in the lead, respectively, and $B$ is the Zeeman field due to ferromagnetism. Equation (S11) produces the imaginary part via the local density of states (LDOS) $\rho_L = -(1/\pi)\text{Im} g_L^r$:

$$[\rho_L(\omega)]_\sigma^{e(h)} = \frac{\theta(2|t_L| - |\omega - \epsilon_\sigma^{e(h)}|)}{|t_L|\pi} \sqrt{1 - \left(\frac{\omega - \epsilon_\sigma^{e(h)}}{2|t_L|}\right)^2}, \tag{S12}$$

which is nonzero within the bandwidth $|(\omega - \epsilon_\sigma^{e(h)})/2|t_z|| < 1$.

Using the above result, we evaluate the self-energy $\Sigma^r = T^\dagger g_L^r T$. Since $T$ is finite between the nearest neighbor sites between the lead and the JJ, the self-energy is nonzero only at sites attached to the lead,

$$\Sigma_{1_S 1_S}^{e(h)}(\omega) = \langle 1_S | T^\dagger | 1_L \rangle \langle 1_L | g^{e(h)}(\omega) | 1_L \rangle \langle 1_L | T | 1_S \rangle, \tag{S13}$$

where $1_L$ and $1_S$ are the adjacent sites on the lead and the JJ, respectively. In terms of the hopping amplitude $\bar{\tau}$ between the lead and the JJ, we have $\langle 1_L | T | 1_S \rangle = -\bar{\tau}\sigma_0$, and thus the self-energy reads

$$\Sigma_{1_S 1_S}^{e(h)} = \begin{pmatrix} |\bar{\tau}|^2 g_{\uparrow\uparrow}^{e(h)} & 0 \\ 0 & |\bar{\tau}|^2 g_{\downarrow\downarrow}^{e(h)} \end{pmatrix}, \tag{S14}$$

with $g_{\uparrow\uparrow}^{e(h)}$ and $g_{\downarrow\downarrow}^{e(h)}$ in Eqs. (S11). Whereas the self-energy term contains both the real and imaginary parts, only the imaginary part gives non-Hermitian effects. The real part only shifts the diagonal entries in $H_{\text{JJ}}$.

To further explore the impact of the non-Hermitian component, we employ the wide-band approximation $|(\omega - \epsilon_\sigma^{e(h)})/(2t_L)| \ll 1$, commonly used in quantum transport [3]. This approximation enables to neglect of the $\omega$-dependence in the Green's function of the lead but still considers $B$ and $\mu_L$ to be large enough to induce different imaginary terms in $g_{\sigma\sigma}^{e(h)}$ for different spins. Now (the imaginary part of) the self-energy reads

$$\Sigma^r(\omega) \approx \begin{pmatrix} \Sigma^e(\omega = 0) & 0 \\ 0 & \Sigma^h(\omega = 0) \end{pmatrix}, \tag{S15}$$

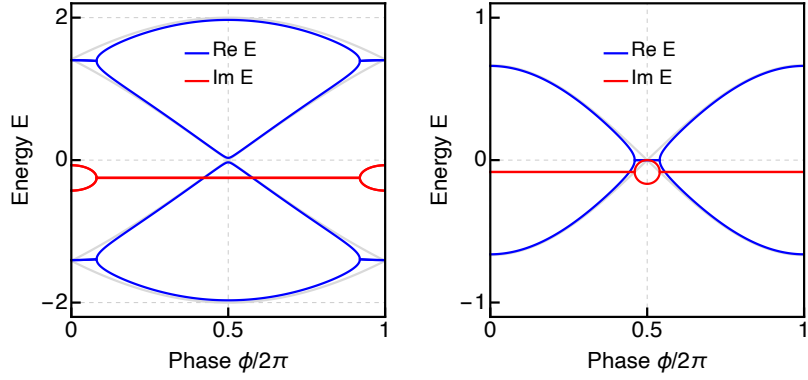

FIG. S1. Left: Re (blue) and Im (red) parts of the eigenvalues given by Eq.(4) of the main text as a function of $\phi$ at $\Gamma_L = 0.5$, $\Gamma_R = 0$. Right: Re (blue) and Im (red) parts of the eigenvalues given by Eq.(6) of the main text as a function of $\phi$ at $\Gamma_\uparrow = 0.5$, $\Gamma_\downarrow = 0$. In both panels, the gray curves correspond to the Hermitian regime. Parameters: $\Delta = 1$, $t = 1$, $\varepsilon_\alpha = 0$.

where

$$\Sigma^{\mathrm{e,h}}(\omega = 0) = -i\Gamma\sigma_0 - i\gamma\sigma_z, \tag{S16}$$

with $\Gamma = (\Gamma_\uparrow + \Gamma_\downarrow)/2$ and $\gamma = (\Gamma_\uparrow - \Gamma_\downarrow)/2$. Here, $\Gamma_\sigma = \pi|\bar{\tau}|^2\rho_\mathrm{L}^\sigma$, where $\rho_\mathrm{L}^\sigma$ is the spin-polarized surface density of states of the lead,

$$\rho_\mathrm{L}^{\uparrow(\downarrow)} = \frac{1}{t_z\pi}\sqrt{1 - \left[\frac{\mu_\mathrm{L} \mp B}{(2t_z)}\right]^2}. \tag{S17}$$

Thus, $\Gamma_\sigma$ is controllable by $B$, $\mu_\mathrm{L}$, $t_z$, and $\bar{\tau}$. The self-energy in Eq. S15 has negative imaginary parts corresponding to decay rates. With the obtained self-energy, the effective Hamiltonian $H_\mathrm{eff}$ in Eq. (S9) describes the JJ coupled to a ferromagnet lead studied in the main text. Similarly, we can obtain the self-energy for a junction coupled to a normal lead, as in the case of the first model discussed in the main text, just by setting $B = 0$ in the above result. The effective Hamiltonian with a zero-frequency self-energy is used in the main text [Eq. (1)] to explore non-Hermitian JJs.

### S1.2. Weak, intermediate, or strong coupling regime

Using the result in the previous section, we compare here the hopping $\bar{\tau}$ between the lead and the JJ with the superconducting pair potential $\Delta$. Depending on the ratio of these energy scales, superconducting heterostructures have three regimes, weak ($\bar{\tau} \ll \Delta$), intermediate ($\bar{\tau} \sim \Delta$), and strong coupling ($\bar{\tau} \gg \Delta$) regions [4].

Here, we should note that $\bar{\tau}$ can be much smaller than $\Gamma_\sigma$: From Eq. (S16), we have

$$\bar{\tau} = \sqrt{\frac{\Gamma_\sigma}{\pi\rho_\mathrm{L}^\sigma}}. \tag{S18}$$

Thus, for the case of $\rho_\mathrm{L}^\sigma = 1$ and $\Gamma_\sigma = 1$, we estimate $\bar{\tau} = 0.56$. Therefore, even when $\Gamma_\sigma$ is comparable with $\Delta = 1$, as in our models in the main text, we may realize the weak coupling regime $\Delta = 1 > \bar{\tau} = 0.56$.

A smaller $\Gamma_\sigma$ easily realizes the weak coupling regime. In Fig. S1, we show the phase-dependent Andreev spectra of the first two models in the main text with a smaller $\Gamma_\sigma$. The spectra clearly indicate exceptional points (EPs), and our conclusion holds even for smaller $\Gamma_\sigma$.

## S2. SYMMETRIES AND NON-HERMITIAN TOPOLOGY OF JOSEPHSON JUNCTIONS

### S2.1. Particle-hole symmetry in the effective Hamiltonian of Josephson Junctions

Here, we discuss particle-hole symmetry (PHS) in the effective Hamiltonian of Josephson Junctions. PHS is a proper symmetry in superconductors, which is defined by

$$\hat{C}\psi_{\alpha i}^\dagger \hat{C}^{-1} = \sum_j \psi_{\alpha j}(U_C^\alpha)^*_{ji}, \quad \hat{C}\psi_{\alpha i}\hat{C}^{-1} = \sum_j \psi_{\alpha j}^\dagger (U_C^\alpha)_{ji}, \tag{S19}$$

where $\psi_{\alpha i}$ represents the Nambu spinor, $\alpha$ is the label distinguishing the JJ ($\alpha=1$) and the lead ($\alpha=2$), $i$ represents the degrees of freedom such as the coordinate and spin, and $U_C^\alpha$ is a unitary matrix. Since PHS is a local symmetry, it does not mix different $\alpha$. For the quadratic Hamiltonian

$$\hat{H} = \sum_{\alpha i,\beta j} \psi_{\alpha i}^\dagger (H)_{\alpha i,\beta j}\psi_{\beta j}, \tag{S20}$$

the above operations reproduce the standard form of PHS:

$$[\hat{H},\hat{C}] = 0 \quad \Leftrightarrow \quad (U_C^\alpha)_{ip}(H^T)_{\alpha p,\beta q}(U_C^{\beta\dagger})_{qj} = -(H)_{\alpha i,\beta j}. \tag{S21}$$

To examine particle-hole symmetry in the effective Hamiltonian, let us consider the retarded Green function of the JJ

$$(G_{\text{JJ}}^{\text{r}})_{ij} = -i\theta(t)\langle\psi_{1i}(t)\psi_{1j}^\dagger + \psi_{1j}^\dagger\psi_{1i}(t)\rangle, \tag{S22}$$

where $\psi_{1i}$ is the Nambu spinor in the JJ. Then, particle-hole symmetry $\hat{C}^{-1}\hat{H}\hat{C} = \hat{H}$ leads to

$$\begin{aligned}
\langle\psi_{1i}(t)\psi_{1j}^\dagger\rangle &= \text{tr}(e^{-\beta\hat{C}^{-1}\hat{H}\hat{C}}e^{i\hat{C}^{-1}\hat{H}\hat{C}t}\psi_{1i}e^{-i\hat{C}^{-1}\hat{H}\hat{C}t}\psi_{1j}^\dagger)/\text{tr}(e^{\beta\hat{H}}) \\
&= \text{tr}(e^{-\beta\hat{H}}e^{i\hat{H}t}\hat{C}\psi_{1i}\hat{C}^{-1}e^{-i\hat{H}t}\hat{C}\psi_{1j}^\dagger\hat{C}^{-1})/\text{tr}(e^{\beta\hat{H}}) \\
&= \left[\sum_{kl}\text{tr}(e^{-\beta\hat{H}}e^{i\hat{H}t}\psi_{1k}^\dagger e^{-i\hat{H}t}\psi_{1l})(U_C^1)_{ki}(U_C^1)^*_{lj}\right]/\text{tr}(e^{\beta\hat{H}}) \\
&= \left[\sum_{kl}(\text{tr}(\psi_{1l}^\dagger e^{i\hat{H}t}\psi_{1k}e^{-i\hat{H}t}e^{-\beta\hat{H}}))^*(U_C^1)_{ki}(U_C^1)^*_{lj}\right]/\text{tr}(e^{\beta\hat{H}}) \\
&= \sum_{kl}(U_C^1)^\dagger_{jl}\langle\psi_{1l}^\dagger\psi_{1k}(t)\rangle^*(U_C^1)_{ki},
\end{aligned} \tag{S23}$$

where we have used the Hermiticity of $\hat{H}$. Thus, we have

$$\begin{aligned}
(G_{\text{JJ}}^{\text{r}})_{ij} &= (U_C^1)^\dagger_{jl}\left[i\theta(t)\langle\psi_{1k}(t)\psi_{1l}^\dagger + \psi_{1l}^\dagger\psi_{1k}(t)\rangle\right]^*(U_C^1)_{ki} \\
&= -(U_C^1)^\dagger_{jl}(G_{\text{JJ}}^{\text{r}})^*_{kl}(U_C^1)_{ki},
\end{aligned} \tag{S24}$$

which leads to particle-hole symmetry in the effective Hamiltonian for the JJ,

$$(U_C^1)^T H_{\text{eff}}^*(-\omega,\phi)(U_C^1)^* = -H_{\text{eff}}(\omega,\phi). \tag{S25}$$

Here, $\phi$ is the phase difference in the JJ, and we make the $\phi$-dependence in $H_{\text{eff}}$ explicit for later convenience. Note that the PHS in the effective Hamiltonian involves complex conjugation, in contrast to the original PHS in Eq.(S21). Following Ref.[5], we call the complex conjugation version of PHS as PHS$^\dagger$. As we show below, PHS$^\dagger$ crucially affects the stability of EPs at zero real energy.

### S2.2. Non-Hermitian topology of the model in Fig. 1(a)

First, we examine the model in Fig. 1 (a) of the main text. Here, we consider an $s$-wave gap function and neglect the spin-orbit interactions both in superconductors and leads. Thus, the system supports spin-rotation symmetry.

Using the spin-rotation symmetry, one can show that the reduced Nambu spinor $\psi_{\alpha i}$ with the form

$$\psi_{\alpha i} = \begin{pmatrix} c_{\alpha \uparrow \underline{i}} \\ c^\dagger_{\alpha \downarrow \underline{i}} \end{pmatrix} \equiv \begin{pmatrix} \psi_{\alpha 1 \underline{i}} \\ \psi_{\alpha 2 \underline{i}} \end{pmatrix} \tag{S26}$$

describes the system, where $\underline{i}$ represents the degrees of freedom other than spin. The Nambu spinor satisfies

$$\hat{S}_{2y} \psi^\dagger_{\alpha 1 \underline{i}} \hat{S}^{-1}_{2y} = \hat{S}_{2y} c^\dagger_{\alpha \uparrow \underline{i}} \hat{S}^{-1}_{2y} = -c^\dagger_{\alpha \downarrow \underline{i}} = -\psi_{\alpha 2 \underline{i}}, \quad \hat{S}_{2y} \psi^\dagger_{\alpha 2 \underline{i}} \hat{S}^{-1}_{2y} = \hat{S}_{2y} c_{\alpha \downarrow \underline{i}} \hat{S}^{-1}_{2y} = c_{\alpha \uparrow \underline{i}} = \psi_{\alpha 1 \underline{i}}, \tag{S27}$$

where $\hat{S}_{2y}$ is the spin rotation by $\pi$ around the $y$-axis, and thus the system supports PHS in Eq.(S19) with

$$(U_C^\alpha)_{ji} = i(\tau_y)_{\tau'\tau} \delta_{\underline{j},\underline{i}}, \quad (i = (\tau, \underline{i}), \ j = (\tau', \underline{j})). \tag{S28}$$

Consequently, the effective Hamiltonian of the JJ hosts the corresponding PHS$^\dagger$ in Eq.(S25, as shown above. We can also confirm this symmetry directly: The model in Fig. 1(a) of the main text is given by

$$H^{(1)}_{\text{eff}}(\phi) = \begin{pmatrix} \varepsilon_L & \Delta & t & 0 \\ \Delta & -\varepsilon_L & 0 & -t \\ t & 0 & \varepsilon_R & \Delta e^{i\phi} \\ 0 & -t & \Delta e^{-i\phi} & -\varepsilon_R \end{pmatrix} + \begin{pmatrix} -i\Gamma_L & 0 & 0 & 0 \\ 0 & -i\Gamma_L & 0 & 0 \\ 0 & 0 & -i\Gamma_R & 0 \\ 0 & 0 & 0 & -i\Gamma_R \end{pmatrix}, \tag{S29}$$

where $\varepsilon_L$, $\varepsilon_R$, $\Delta$, and $\phi$ are real constants. This model has PHS$^\dagger$ with

$$U_C = \begin{pmatrix} 0 & -1 & 0 & 0 \\ 1 & 0 & 0 & 0 \\ 0 & 0 & 0 & -1 \\ 0 & 0 & 1 & 0 \end{pmatrix}. \tag{S30}$$

Because $U_C$ satisfies $U_C U_C^* = -1$, this symmetry defines class C$^\dagger$ in the symmetry classification in Ref.[5]. This particular type of PHS$^\dagger$ explains why the level crossing at $\phi = \pi$ does not evolve into EPs: According to a general theory in Ref.[6], the point-gap topology intrinsic to non-Hermitian systems ensures the robust existence of EPs. For an EP in the space of $\phi$, the relevant point-gap topological number is 0-dimensional. Moreover, since the level crossing at zero energy is invariant under PHS$^\dagger$, and thus, the point-gap topological number should respect PHS$^\dagger$. However, any class C$^\dagger$ system does not have a 0-dimensional point-gap topological number so the level crossing at zero energy does not evolve into robust EPs.

If we assume $\varepsilon_L = \varepsilon_R$, as in our calculations in Fig.2, this model also has an accidental parity-time (PT) symmetry,

$$PT \left[ H^{(1)}_{\text{eff}}(\phi) + i\frac{\Gamma_L + \Gamma_R}{2} 1 \right]^* T^\dagger P^\dagger = H^{(1)}_{\text{eff}}(\phi) + i\frac{\Gamma_L + \Gamma_R}{2} 1, \tag{S31}$$

where the unitary matrices for time-reversal $T$ and parity $P$ are given by

$$T = \begin{pmatrix} e^{i\phi/2} & 0 & 0 & 0 \\ 0 & e^{-i\phi/2} & 0 & 0 \\ 0 & 0 & e^{i\phi/2} & 0 \\ 0 & 0 & 0 & e^{-i\phi/2} \end{pmatrix}, \quad P = \begin{pmatrix} 0 & 0 & 1 & 0 \\ 0 & 0 & 0 & 1 \\ 1 & 0 & 0 & 0 \\ 0 & 1 & 0 & 0 \end{pmatrix}, \tag{S32}$$

and satisfy $PT(PT)^* = 1$. The PT symmetry enables stable EPs with the energy $E$, which imaginary part is pinned at $\text{Im}E = -(\Gamma_L + \Gamma_R)/2$. Note that EPs in Fig. 2(a) have such an imaginary part of the energy. As shown in Ref.[6], the PT-symmetry enables a 0-dimensional $\mathbb{Z}_2$ point-gap topological number, ensuring the stability of EPs in the one-dimensional $\phi$ space. Therefore, the analysis carried out here help understanding the absence and presence of EPs at zero and finite energies, respectively, in the Section "NH JJs with superconductors coupled to different normal leads" of the main text.

### S2.3. Non-Hermitian topology of the model in Fig. 1(b)

In Fig.1 (b) of the main text, we consider a ferromagnetic lead that breaks spin-rotation symmetry. Therefore, the full Nambu spinor

$$\psi_{\alpha i} = \begin{pmatrix} c_{\alpha\uparrow\underline{i}} \\ c_{\alpha\downarrow\underline{i}} \\ c^\dagger_{\alpha\uparrow\underline{i}} \\ c^\dagger_{\alpha\downarrow\underline{i}} \end{pmatrix} \equiv \begin{pmatrix} \psi_{\alpha 1\uparrow\underline{i}} \\ \psi_{\alpha 1\downarrow\underline{i}} \\ \psi_{\alpha 2\uparrow\underline{i}} \\ \psi_{\alpha 2\downarrow\underline{i}} \end{pmatrix} \qquad (S33)$$

is necessary to describe the system. Since we have the identity

$$\psi^\dagger_{\alpha\tau\sigma\underline{i}} = \sum_{\tau'} \psi_{\alpha\tau'\sigma\underline{i}} (\tau_x)_{\tau'\tau}, \qquad (S34)$$

the Hamiltonian should have PHS in Eq.(S19) with $\hat{C} = 1$ and

$$(U^\alpha_C)_{ji} = (\tau_x)_{\tau'\tau}(\sigma_0)_{\sigma'\sigma}\delta_{\underline{j},\underline{i}}, \quad (j=(\tau',\sigma',\underline{j}),\, i=(\tau,\sigma,\underline{i})). \qquad (S35)$$

Therefore, the effective Hamiltonian of the JJ obeys PHS$^\dagger$ in Eq.(S25). In the present case, PHS$^\dagger$ defines class D$^\dagger$ in the symmetry classification since $U^\alpha_C$ satisfies $U^\alpha_C(U^\alpha_C)^* = 1$.

In contrast to class C$^\dagger$ in the previous model, class D$^\dagger$ supports a 0-dimensional $\mathbb{Z}_2$ point-gap topological number [6]. Actually, we can confirm that EPs at zero real energy in Fig. 3 (a) and Figs. 4 (a-c) have the non-trivial $\mathbb{Z}_2$ number. Because PHS$^\dagger$ in class D$^\dagger$ is intact under any perturbation in superconductors, these EPs are robust unless the superconducting gap closes or they are pair-annihilated. These EPs are discussed in the section on "NH JJs with a middle N region coupled to a ferromagnet lead" of the main text.

## S3. ANDREEV BOUND STATES IN THE HERMITIAN REGIME

In this section, we compare the Andreev bound states (ABSs) of the models in Eqs. (3), and (5) in the absence of non-Hermiticity with those commonly used in the literature.

### S3.1. Model in Eq. (4) in the Hermitian regime

The eigenvalues for the first model [Eq. (4)] without non-Hermiticity are given by

$$E_j = \pm\sqrt{\delta^2 \pm 2t\Delta |\sin(\phi/2)|}, \qquad (S36)$$

where $\delta = t^2 + \Delta^2$, and $j = 1, 2, 3, 4$ is the index for the different eigenvalues. Thus, in-gap ABSs in the Hermitian regime are given by

$$E_{2,3} = \pm\sqrt{\delta^2 - 2t\Delta |\sin(\phi/2)|}, \qquad (S37)$$

for $t\Delta > 0$. When $|t| = |\Delta|$, the in-gap ABSs show a level crossing at zero energy at $\phi = \pi$, as illustrated in gray curves in Fig.2(a) of the main text.

### S3.2. Model in Eq. (6) in the Hermitian regime

The second model in Eq. (6) hosts in-gap ABSs with the eigenvalues in Eq.(6) in the main text. Without non-Hermiticity, the eigenvalues read

$$E_\pm = \pm\frac{\Delta}{A}\sqrt{C + 2t^4 A \cos(\phi)} \qquad (S38)$$

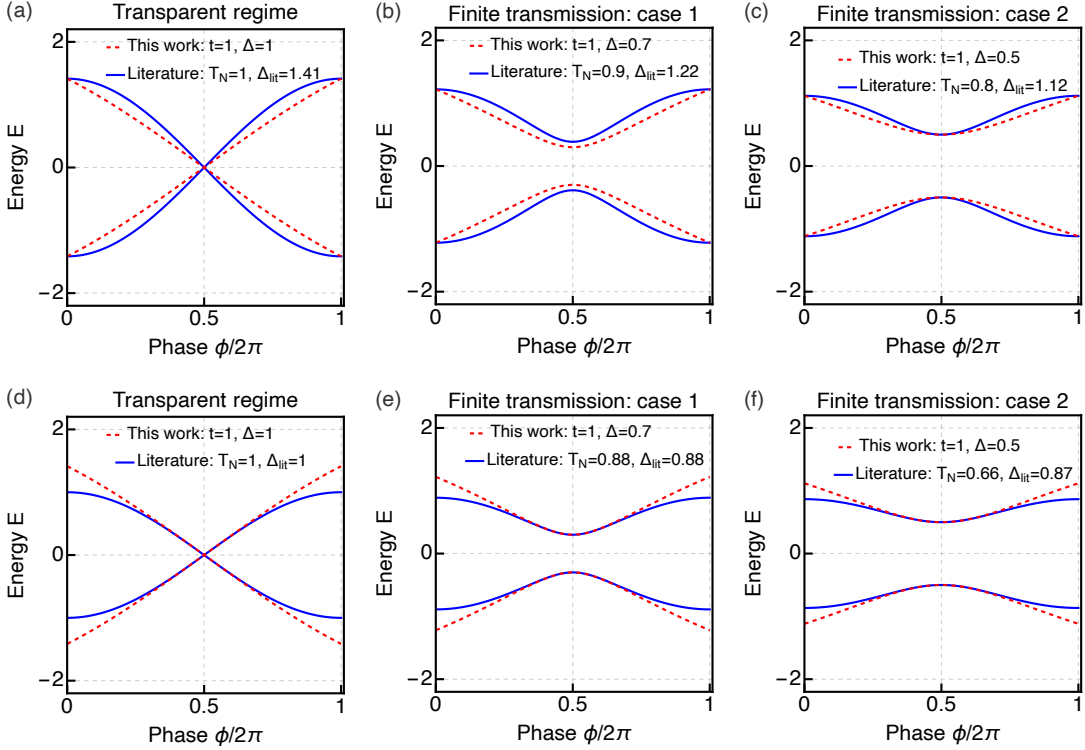

FIG. S2. ABSs as a function of $\phi$ in the Hermitian regime of our first model given by Eq. (S36) (red dashed curves) and using the ABSs from the literature given by Eq. (S40) (blue solid curves). The left, middle, and right columns correspond to distinct parameters, capturing the transparent and finite transmission regimes. (a-c) the ABSs in Eq. (S36) and Eq. (S40) show qualitatively same behaviors. (d-f) The ABSs in Eq. (S36) and Eq. (S40) exhibit exactly the same $\phi$-dependence near $\phi \sim \pi$. Our models, therefore, reproduce the conventional description for ABSs in Hermitian cases.

with $A = 4t^4 + \Delta^4 + 4\Delta^2 t^2$ and $C = 2t^4[4t^4 + (4t^2)\Delta^2 + \Delta^4]$. Since we have $2t^4 A = C$ in the Hermitian case, the eigenvalues are recast into

$$E_\pm = \pm \frac{\Delta\sqrt{C}}{A}\sqrt{1+\cos(\phi)} = \pm \frac{\Delta\sqrt{2C}}{A}\cos(\phi/2). \tag{S39}$$

Thus, they show a level crossing at $\phi = \pi$. In the main text, we illustrate these ABSs in the Hermitian regime in the gray curves of Fig. 3 (a).

### S3.3. Comparison with the standard Hermitian ABSs in the literature

Here, we compare the above ABSs [Eqs. (3) and (5) of the main text] in the Hermitian regime with the standard ABSs in literature. The well-known expressions for the ABSs are commonly derived within the Andreev approximation, where the energy scales are much smaller than the Fermi energy. Here, we focus on the expression for short JJs, as in our models. The standard ABSs in short JJs read [7–9]

$$E_\pm^{\text{lit}} = \pm \Delta_{\text{lit}} \sqrt{1 - T_{\text{N}} \sin^2(\phi/2)}, \tag{S40}$$

where $\Delta_{\text{lit}}$ is the superconducting gap, and $T_{\text{N}}$ is the normal transmission. For $T_{\text{N}} = 1$, it holds $E_\pm^{\text{lit}} = \pm\Delta_{\text{lit}}\cos(\phi/2)$; thus the ABSs show a level crossing with zero energy at $\phi = \pi$.

*Comparison with model in Eq. (4)*

The ABSs in Eq. (S36) and (S40) show qualitatively the same behavior as functions of $\phi$. See Fig. S2 (a-c). In both cases, they have the maximal (minimal) gap at $\phi = 0$ ($\phi = \pi$). The exact $\phi$-dependence is not the same between

them, but it comes from the difference in the bulk degrees of freedom. Whereas the derivation of Eq. (S40) considers semi-infinite superconductors [7–9], Eq. (S36) considers single-site superconductors. Thus, we can expect the exact coincidence only when the ABSs rarely mix with the bulk states: For $\phi \sim \pi$, where the ABSs are distinct from the bulk states, Eqs. (S36) and (S40) give

$$\begin{aligned} E_{2,3} &= \pm\sqrt{(t-\Delta)^2 + \frac{1}{4}t\Delta(\phi-\pi)^2}\,, \\ E_\pm^{\text{lit}} &= \pm\sqrt{(1-T_N)\Delta_{\text{lit}}^2 + \frac{1}{4}T_N\Delta_{\text{lit}}^2(\phi-\pi)^2}\,, \end{aligned} \quad \text{(S41)}$$

and thus, they coincide exactly by the identification $(t-\Delta)^2 = (1-T_N)\Delta_{\text{lit}}^2$ and $t\Delta \sim T_N\Delta_{\text{lit}}^2$. In contrast, they show a deviation for $\phi \sim 0$, since the ABSs are much closer to the bulk states in this situation. See Fig. S2 (d-e).

*Comparison with model in Eq. (6)*

The ABSs in Eq. (S39) exhibit the exactly same $\phi$-dependence in Eq. (S40) with $T_N = 1$ and $(\Delta\sqrt{2C}/A) = \Delta_{\text{lit}}$.

### S4. ANALYTICAL EXPRESSIONS FOR THE SUPERCURRENTS OF THE EFFECTIVE MODELS

This section presents analytical expressions of the supercurrents associated with the ABSs of the effective models in Eqs. (3), and (5) of the main text. The ABSs for these models are given by Eqs. (4) and (6).

For the model in Eq. (3), the total supercurrent is given by the two positive ABSs, namely, $E_{1,2}$, where $E_{1(2)}$ is the higher (lower) positive ABS in Eq. (4) of the main text. From Eq. (2) of the main text, each of these ABSs carries the supercurrent of $I_{1(2)} = (-e/\hbar)dE_{1(2)}/d\phi$, which is then given by

$$\begin{aligned} I_1(\phi) &= -\frac{e}{\hbar}\frac{dE_1}{d\phi} = -\frac{t^2\Delta\sin(\phi)}{4\sqrt{t^2\sin^2(\phi/2)-\gamma^2}\sqrt{\delta^2-\gamma^2+2\Delta\sqrt{t^2\sin^2(\phi/2)-\gamma^2}}}\,, \\ I_2(\phi) &= -\frac{e}{\hbar}\frac{dE_2}{d\phi} = \frac{t^2\Delta\sin(\phi)}{4\sqrt{t^2\sin^2(\phi/2)-\gamma^2}\sqrt{\delta^2-\gamma^2-2\Delta\sqrt{t^2\sin^2(\phi/2)-\gamma^2}}}\,, \end{aligned} \quad \text{(S42)}$$

where $\delta = t^2 + \Delta^2$ and $\gamma = (\Gamma_L - \Gamma_R)/2$. As discussed in the main text, EPs appear at finite real energies when $t^2\sin^2(\phi/2) = \gamma^2$, which then implies that both supercurrents diverge at EPs but with opposite sign. Thus, at EPs the divergent profile of the total supercurrent, the sum of both $I_1^+$ and $I_1^-$, is suppressed, giving rise to a smooth behavior across EPs. Between EPs, when $t^2\sin^2(\phi/2) < \gamma^2$, the current exhibits both real and imaginary parts due to finite phase dependent real and imaginary ABSs, as seen in Fig. 2 of the main text.

For the second model in Eq.(5), we have two ABSs around zero energy. See Eq. (6) in the main text. Again, only the positive ABS contributes to the total supercurrent, as shown in Eq. (2) of the main text. Thus, the total supercurrent in the second model is given by

$$I^+(\phi) = -\frac{e}{\hbar}\frac{dE_+}{d\phi} = +\Delta t^4 \frac{\sin(\phi)}{\sqrt{C+2t^4A\cos(\phi)}} \quad \text{(S43)}$$

where $C = 4t^4[2t^4+(2t^2-2\gamma^2+\Gamma^2)\Delta^2]+2[(t^2-\gamma^2)^2+\Gamma^2(\Gamma^2-2\gamma^2)]\Delta^4-\gamma^2\Delta^6$ and $A = 4t^4+\Delta^4+2\Delta^2\left(-\gamma^2+\Gamma^2+2t^2\right)$. Here, EPs form at zero-real energy when $C + 2t^4A\cos(\phi) = 0$, implying that the supercurrent diverges at these EPs. Unlike the supercurrent in the first model, no suppression occurs for $I^+$ because no other energy level contributes to the total supercurrent. Thus, EPs enhance the supercurrent as we indeed observe in Fig. 3(c) of the main text.

### S5. COMPARING THE SUPERCURRENT FORMULAS USED IN NON-HERMITIAN JOSEPHSON JUNCTIONS: ENHANCED SUPERCURRENT BY EXCEPTIONAL POINTS

Recently, Shen and collaborators derived a formula for the supercurrent in a non-Hermitian JJ [10] given by,

$$I(\phi) = -\frac{e}{\hbar\pi}\frac{d}{d\phi}\text{Im}\text{Tr}(H_{\text{eff}}(\phi)\ln H_{\text{eff}}(\phi))\,, \quad \text{(S44)}$$

where $H_{\text{eff}}(\phi)$ is the effective Bogoliubov-de Gennes Hamiltonian defined by the retarded Green's function of the junction

$$G^r(\omega) = \frac{1}{\omega - H_{\text{eff}}(\phi)}, \tag{S45}$$

and $\phi$ is the phase difference between the left and right superconductors of the junction. Below, we compare this formula [Eq. (S44)] with the semi-classical current formula used in the main text [Eq. (2)],

$$I_0(\phi) = -\frac{e}{\hbar}\sum_{n\geq 0}\left[\frac{d\text{Re}E_n(\phi)}{d\phi}\right], \tag{S46}$$

where $E_n(\phi)$ represents the eigenvalues of $H_{\text{eff}}(\phi)$ and we take the summation $n$ for all $E_n(\phi)$s' with $\text{Re}E_n(\phi) \geq 0$. In particular, we show that both formulas support the idea that an exceptional point intrinsically has a divergent contribution to the Josephson current.

For this purpose, we first rewrite the formula in Eq. (S44) in terms of the eigenvalues $E_n(\phi)$ of $H_{\text{eff}}$. Using a unitary transformation, $H_{\text{eff}}$ can be an upper triangular form, of which the diagonal components are eigenvalues of $H_{\text{eff}}$. Thus, Eq. (S44) is recast into

$$I(\phi) = -\frac{e}{\hbar\pi}\frac{d}{d\phi}\text{Im}\left(\sum_n E_n(\phi)\ln E_n(\phi)\right), \tag{S47}$$

where we take the summation $n$ for all possible eigenvalues $E_n(\phi)$. Since $H_{\text{eff}}(\phi)$ has the following particle-hole symmetry,

$$CH_{\text{eff}}^*(\phi)C^\dagger = -H_{\text{eff}}(\phi), \tag{S48}$$

with a unitary matrix $C$, an eigenvalue $E_n$ forms a pair with another eigenvalue $-E_n^*$. Therefore, we can further rewrite the above equation as

$$I(\phi) = -\frac{e}{\hbar\pi}\frac{d}{d\phi}\text{Im}\sum_{n\geq 0}\left(E_n(\phi)\ln E_n(\phi) - E_n^*(\phi)\ln(-E_n^*(\phi))\right)$$

$$= -\frac{e}{\hbar\pi}\frac{d}{d\phi}\text{Im}\sum_{n\geq 0}\left(\text{Re}E_n(\phi)\ln(-E_n(\phi)/E_n^*(\phi)) + i\text{Im}E_n(\phi)\ln(-|E_n(\phi)|^2)\right)$$

$$= -\frac{e}{\hbar\pi}\frac{d}{d\phi}\sum_{n\geq 0}\left(\text{Re}E_n(\phi)\text{Im}[\ln(-E_n(\phi)/E_n^*(\phi))] + \text{Im}E_n(\phi)\ln|E_n(\phi)|^2\right), \tag{S49}$$

where now we perform the summation $n$ for all $E_n$s' with $\text{Re}E_n(\phi) \geq 0$. Then, using the relation,

$$\ln(-E_n(\phi)/E_n^*(\phi)) = 2i\arctan(-\text{Re}E_n(\phi)/\text{Im}E_n(\phi)), \tag{S50}$$

we have

$$I(\phi) = -\frac{e}{\hbar\pi}\frac{d}{d\phi}\sum_{n\geq 0}\left(2\text{Re}E_n(\phi)\arctan\left(\frac{-\text{Re}E_n(\phi)}{\text{Im}E_n(\phi)}\right) + \text{Im}E_n(\phi)\ln|E_n(\phi)|^2\right). \tag{S51}$$

To examine the influence of an exceptional point on the Josephson current $I(\phi)$, let us consider an eigenenergy $E_0(\phi)$ given by

$$E_0 = \sqrt{s(\phi)} - i\Gamma, \tag{S52}$$

where $s(\phi)$ is a real function of $\phi$. When $s(\phi) = 0$, $E_0(\phi)$ and $-E_n^*(\phi)$ are degenerate and form an exceptional point at zero real energy, as we shown in Fig. 3(a) of the main text. Then, for $s(\phi) \geq 0$, $E_0$ contributes to the supercurrent [Eq. (S51)] as [11, 12]

$$I(\phi)|_{E_0} = -\frac{e}{\hbar}\left[\frac{d\sqrt{s(\phi)}}{d\phi}\right]\left[\frac{2}{\pi}\arctan\left(\frac{\sqrt{s(\phi)}}{\Gamma}\right)\right]$$

$$= I_0(\phi)|_{E_0}\left[\frac{2}{\pi}\arctan\left(\frac{\sqrt{s(\phi)}}{\Gamma}\right)\right], \tag{S53}$$

where $I_0(\phi)|_{E_0}$ is the contribution of $E_0$ to the semi-classical current $I_0(\phi) = (-e/\hbar)[d\sqrt{s(\phi)}/d\phi]$. In the derivation of Eq. (S53), we have used the relation

$$2\frac{d}{d\phi}\arctan\left(\frac{\sqrt{s(\phi)}}{\Gamma}\right) = \Gamma\frac{d}{d\phi}\ln\bigl(s(\phi) + \Gamma^2\bigr) = \frac{2\Gamma}{s^2(\phi) + \Gamma^2}\frac{d\sqrt{s(\phi)}}{d\phi}. \tag{S54}$$

Now, we argue that there exists an intrinsic contribution of the exceptional point to the Josephson current. Since $\Gamma$ is not related to the existence of the exceptional point, we should take $\Gamma = 0$ to reveal the intrinsic contribution. Then, we find from Eq. (S53) that the Josephson current coincides with the semi-classical formula

$$I(\phi)|_{E_0} = I_0(\phi)|_{E_0}, \tag{S55}$$

which diverges at the exceptional point. Thus, the intrinsic contribution of the exceptional point to the Josephson current is divergent.

In actual experimental situations, we usually have a nonzero $\Gamma$. Since the arctangent factor in Eq.(S53) gives a cut-off, it obscures the divergent current due to the exceptional point [13]. Still, the exceptional point enhances the Josephson current for small $\Gamma$. From Eq.(S53), we can estimate the peak of the current near the exceptional point as

$$I(\phi)|_{E_0} \sim \frac{e}{\hbar}\frac{s'(\phi)}{\Gamma}\bigg|_{s(\phi)=0}. \tag{S56}$$

Whereas Shen et al. in [10] do not specify the origin of the enhancement of the Josephson current in Fig. 2(a) of their paper, it can be associated with the exceptional point in Fig. S1(a) of their work. Thus, contrary to their claim, our argument is consistent with their numerical result.


\end{document}